\documentclass{ws-ijmpe}
\usepackage{wrapfig}

\begin{document}

\markboth{Xiaoyan Lin (for the STAR Collaboration)}{Measurement of
Non-Photonic Electron Angular Correlations with Charged Hadrons}

\catchline{}{}{}{}{}

\title{MEASUREMENT OF NON-PHOTONIC ELECTRON ANGULAR CORRELATIONS WITH CHARGED HADRONS}

\author{\footnotesize XIAOYAN LIN (for the STAR Collaboration)}

\address{Institute of Particle, Central China Normal University\\Wuhan, Hubei 430079, China\\
linxy@iopp.ccnu.edu.cn}

\maketitle


\begin{abstract}
Correlations of non-photonic electrons with charged hadrons are
studied using the PYTHIA Monte Carlo event generator in p+p
collisions at $\sqrt{s_{NN}} = 200$ GeV. We propose experimental
methods to estimate the relative contributions to non-photonic
electrons from charm and bottom decays. We present the measurement
of azimuthal correlations between non-photonic electrons and charged
hadrons in p+p collisions at $\sqrt{s_{NN}} = 200$ GeV from STAR.
The results are compared to PYTHIA simulations to estimate the
relative contributions of $D$ and $B$ meson semi-leptonic decays to
non-photonic electrons. Our preliminary result indicates that at
$p_{T} \sim 4.0 - 6.0$ GeV/c the measured $B$ contribution to
non-photonic electrons is comparable to $D$ contribution, and is
consistent with the FONLL calculation.
\end{abstract}

\section{Introduction}

High transverse momentum particles in central Au+Au collisions are
found to be significantly suppressed compared to that from p+p
collisions.\cite{adams,adler} This is consistent with the picture
that high energy partons are quenched when propagating through the
dense medium created in central nucleus-nucleus
collisions;\cite{wang} and this suppression is thus a probe for the
Quark Gluon Plasma. Light hadron suppression at RHIC has been
described by a gluon radiation energy loss
mechanism.\cite{adil,vitev} Recent RHIC data\cite{abelev,adler2}
show that the suppression of non-photonic electrons (electrons from
heavy quark decays) has almost the same magnitude as that of light
hadrons in central Au+Au collisions. This implies that heavy quarks
may lose a substantial amount of energy, which cannot be explained
by current theoretical predictions based on gluon radiation as the
dominant mechanism for energy loss.\cite{armesto,djordjevic} Heavy
quark would lose much less energy because the gluon radiation is
suppressed at small angles due to the mass of heavy quarks (dead
cone effect\cite{doks}). Recent calculations including both gluon
radiation and collisional energy loss increase the heavy quark
energy loss.\cite{mustafa,adil2} In both scenarios electrons from
charm decays are always more suppressed than those from bottom.
However, theoretical calculations contain large uncertainties with
respect to the relative contributions to non-photonic electrons from
charm and bottom quarks. Another important tool to study the Quark
Gluon Plasma properties is the angular anisotropy $v_{2}$.
Observation of heavy quark hydrodynamic flow is a unique probe for
the early stage of a partonic phase.\cite{greco} A nonzero
non-photonic electron $v_{2}$ has been measured for $p_{T} < 2.0$
GeV/c, while at higher $p_{T}$ (although uncertainties are still
large) the $v_{2}$ is observed to decrease with $p_{T}$.\cite{adare}
The same $D$ and $B$ hadron $v_{2}$ can lead to very different
non-photonic electron $v_{2}$ due to the different decay kinematics
between $D$ and $B$ hadrons.\cite{zhang} The quantitative
understanding of the non-photonic electron suppression at high
$p_{T}$ and $v_{2}$ requires the knowledge of the relative charm and
bottom contributions to non-photonic electrons.

In this paper, correlations of non-photonic electrons with charged
hadrons are studied using the PYTHIA event generator in p+p
collisions at $\sqrt{s_{NN}} = 200$ GeV. We develop an innovative
method which uses the azimuthal correlations between non-photonic
electrons and charged hadrons to estimate the relative $D$ and $B$
contributions to non-photonic electrons. We present the preliminary
results on the measurement of azimuthal correlations between
high-$p_{T}$ non-photonic electrons and charged hadrons in p+p
collisions at $200$ GeV from STAR. We will use comparisons of the
experimental results with PYTHIA simulations to estimate $B$ and $D$
decay contributions to non-photonic electrons as a function of
$p_{T}$ for $p_{T} > 2.5 $ GeV/c.

\section{PYTHIA Simulation}
We have studied the correlations between non-photonic electrons and
inclusive charged hadrons in p+p collsions at $\sqrt{s_{NN}} = 200 $
GeV from PYTHIA $v6.22$.\cite{pythia} The PYTHIA parameters have
been tuned in order to simultaneously describe the STAR measurements
of the $D$ meson $p_{T}$ shape\cite{star3,tai} and of the single
non-photonic electron $p_{T}$ distribution.\cite{abelev} The
parameters for charm quarks are: PARP(67) $=$ 4 (factor multiplied
to $Q^2$) , $<k_t> = 1.5$ GeV/c, $m_c = 1.25$ GeV/$c^2$,
$K_{factor}$ = 3.5, MSTP(33) $=$ 1 (inclusion of $K$ factors),
MSTP(32) $=$ 4 ($Q^2$ scale), CTEQ5L PDF and $\delta$ fragmentation
function. The parameters for bottom quarks are the same as for charm
quarks except $m_b = 4.8$ GeV/$c^2$. The details of PYTHIA parameter
setup are discussed in Reference.\cite{Lin}

\begin{figure}[th]
\centerline{\psfig{file=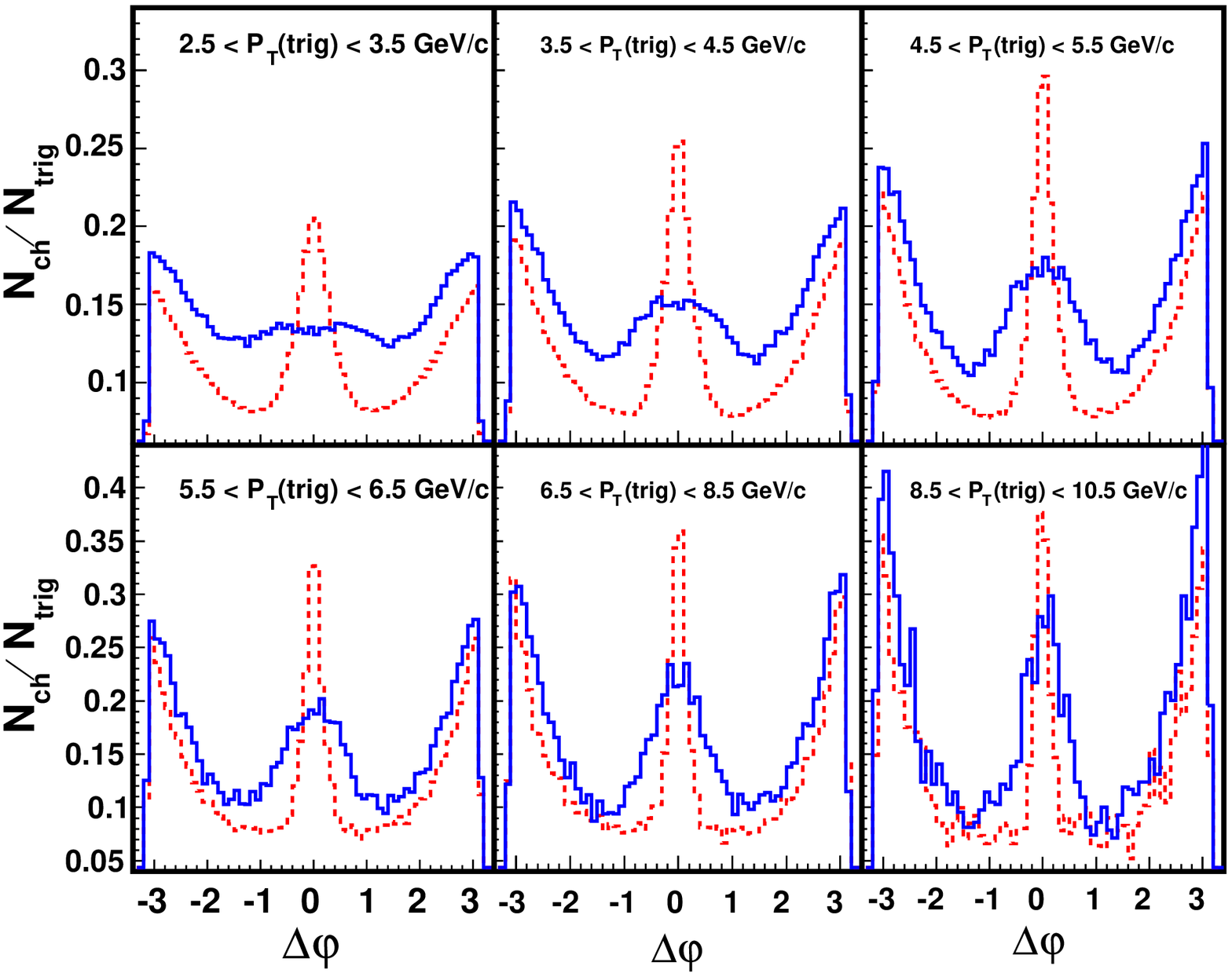,width=6.5cm}
\psfig{file=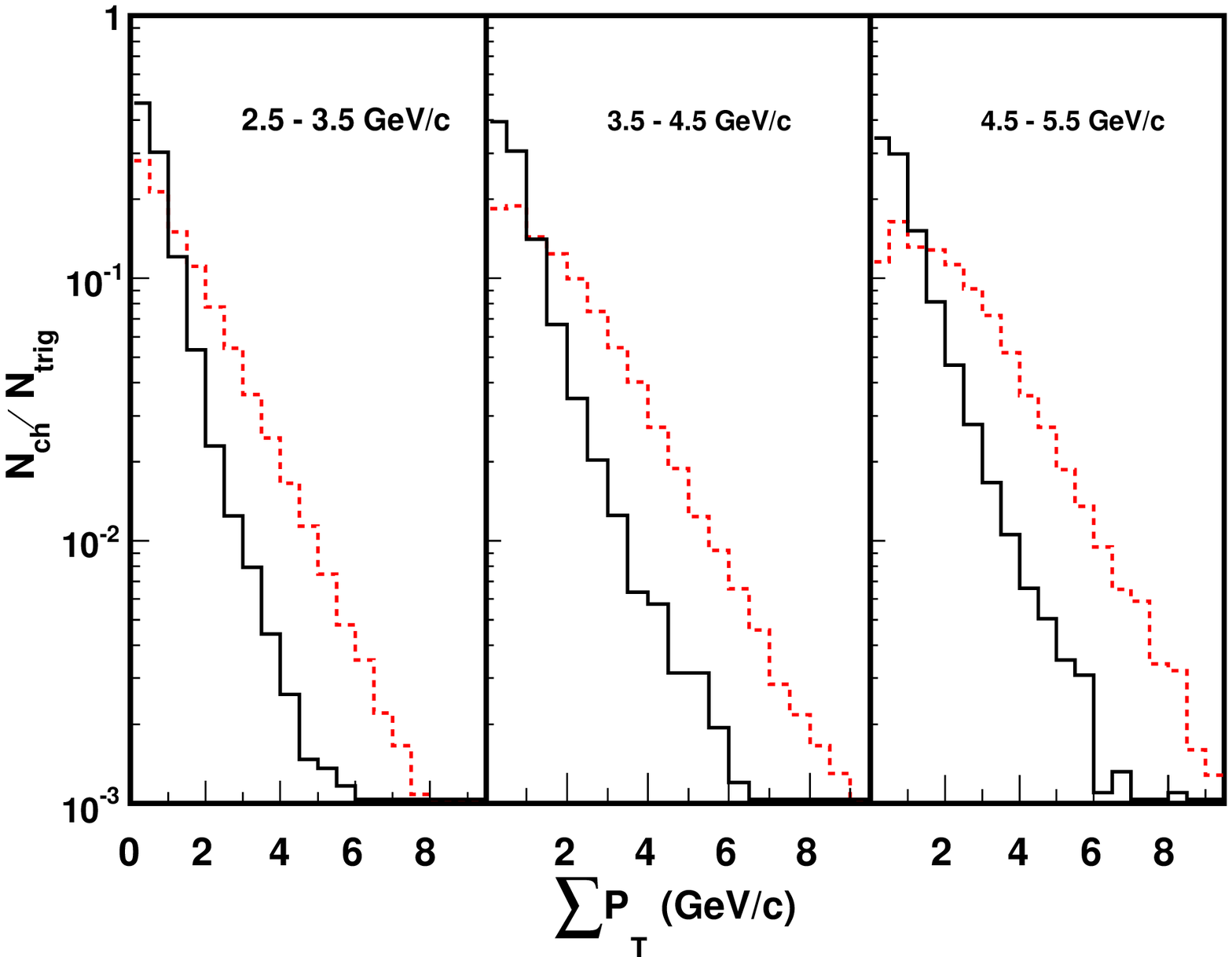,width=6.5cm}} \vspace*{8pt}
\vspace{-0.3cm} \caption{(color online) Left: PYTHIA simulations of
angular correlations between non-photonic electrons and charged
hadrons with six electron trigger $p_{T}$ cuts and associated hadron
$p_{T} > 0.3 $ GeV/c. Solid lines show electrons from $B$ meson
decays. Dashed lines show electron from $D$ meson decays. Right:
PYTHIA simulations of summed $p_{T}$ distributions of charged
hadrons around triggered non-photonic electrons with three trigger
$p_{T}$ cuts. Solid lines show electrons from $B$ meson decays.
Dashed lines show electrons from $D$
decays.}\label{fig1}\vspace{-0.3cm}
\end{figure}

\begin{figure}[th]
\centerline{\psfig{file=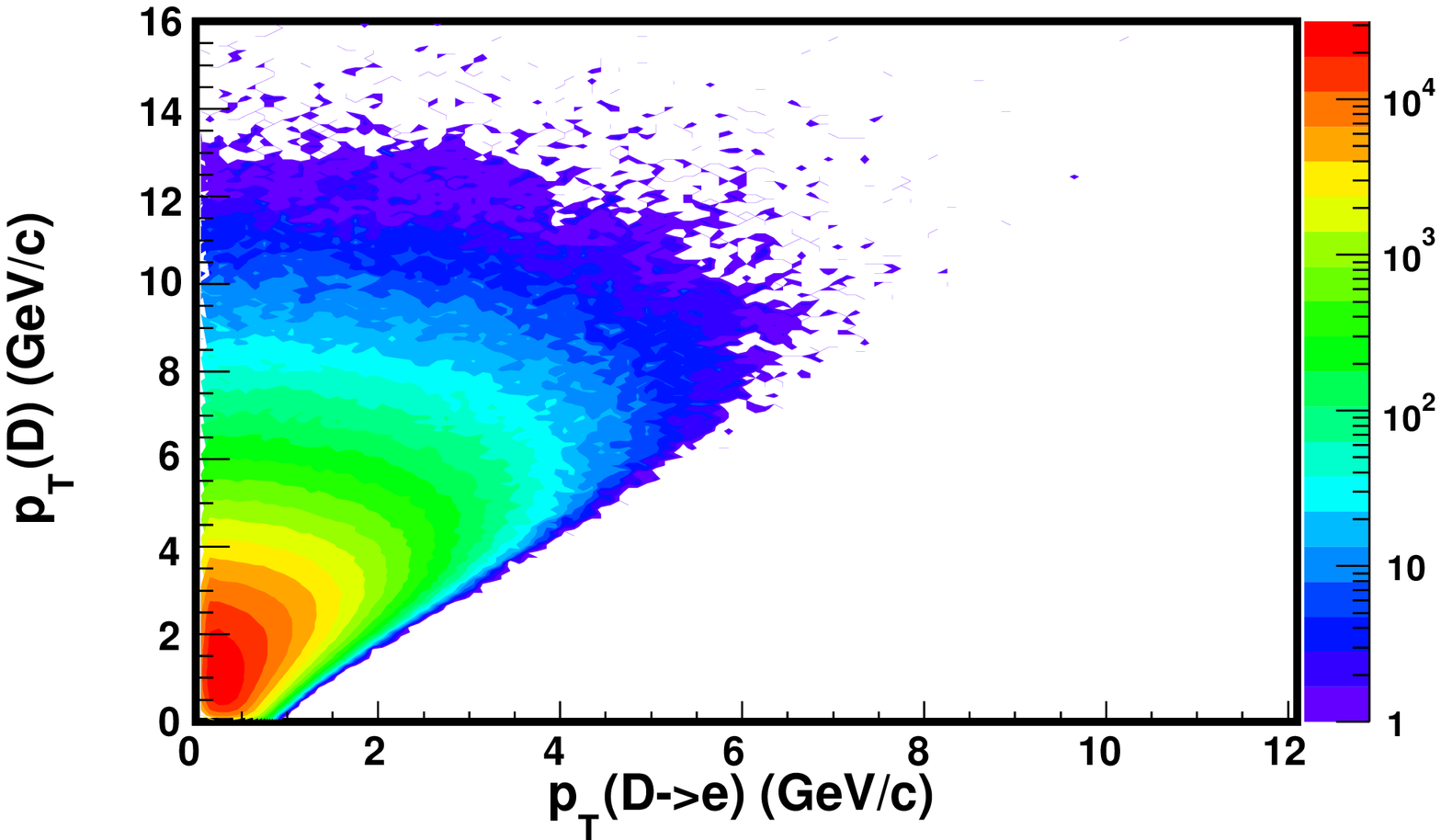,width=6.0cm}
\psfig{file=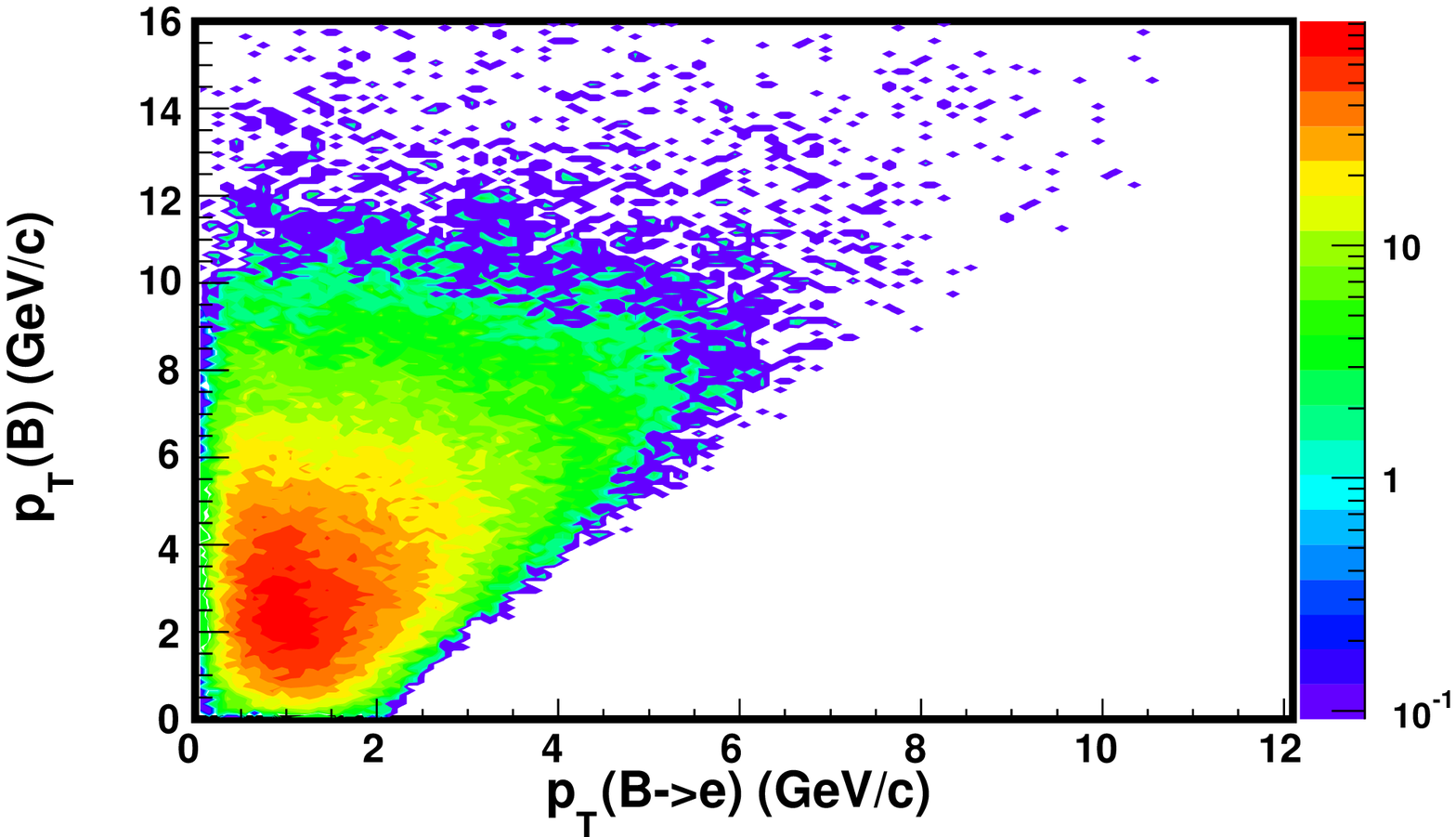,width=6.0cm}} \vspace*{8pt}
\vspace{-0.3cm} \caption{(color online) The distributions of
electron $p_{T}$ versus its parent $p_{T}$ (left for $D$ decays and
right for $B$ decays).}\label{fig2} \vspace{-0.3cm}
\end{figure}

\begin{figure}[th]
\centerline{\psfig{file=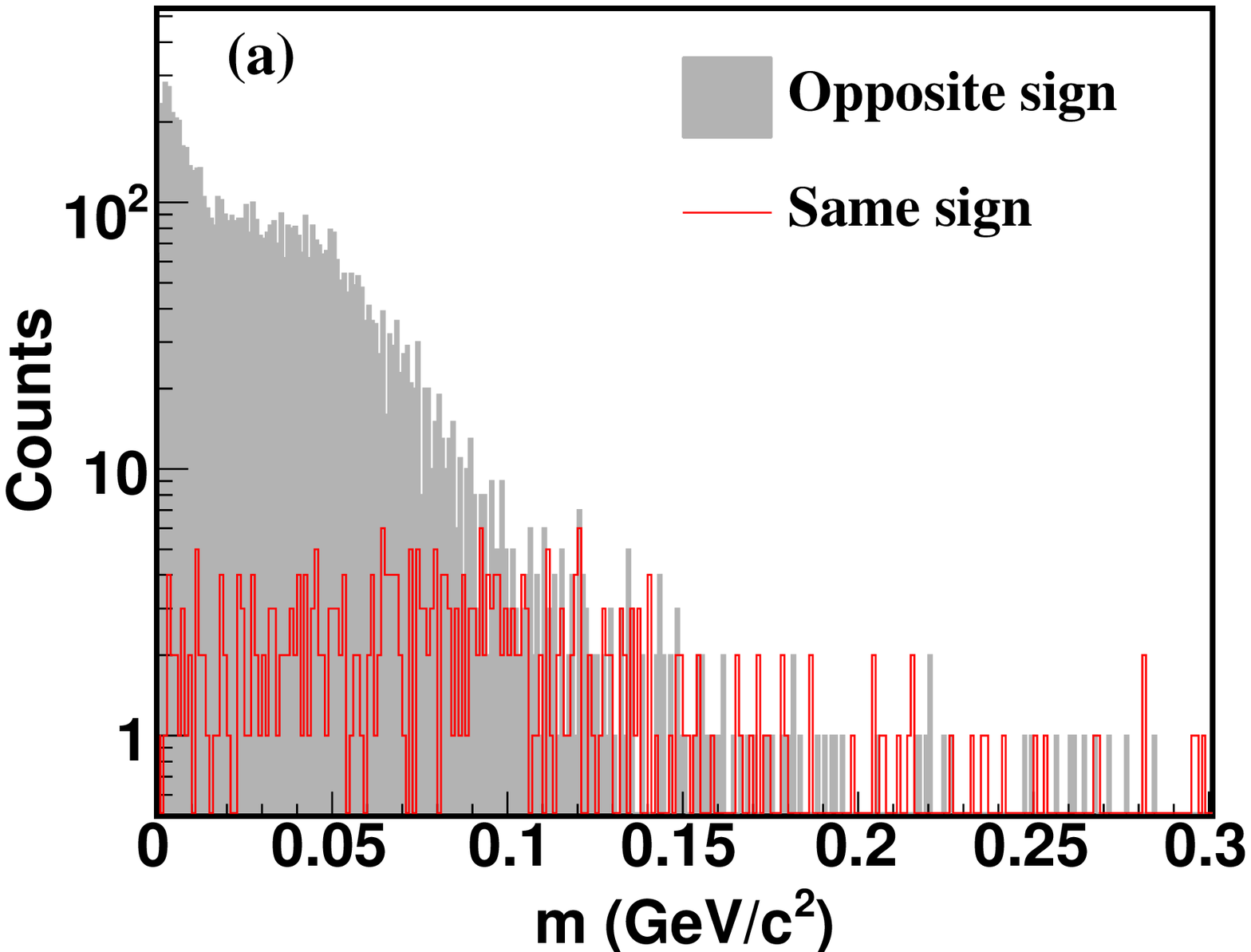,width=6.5cm}
\psfig{file=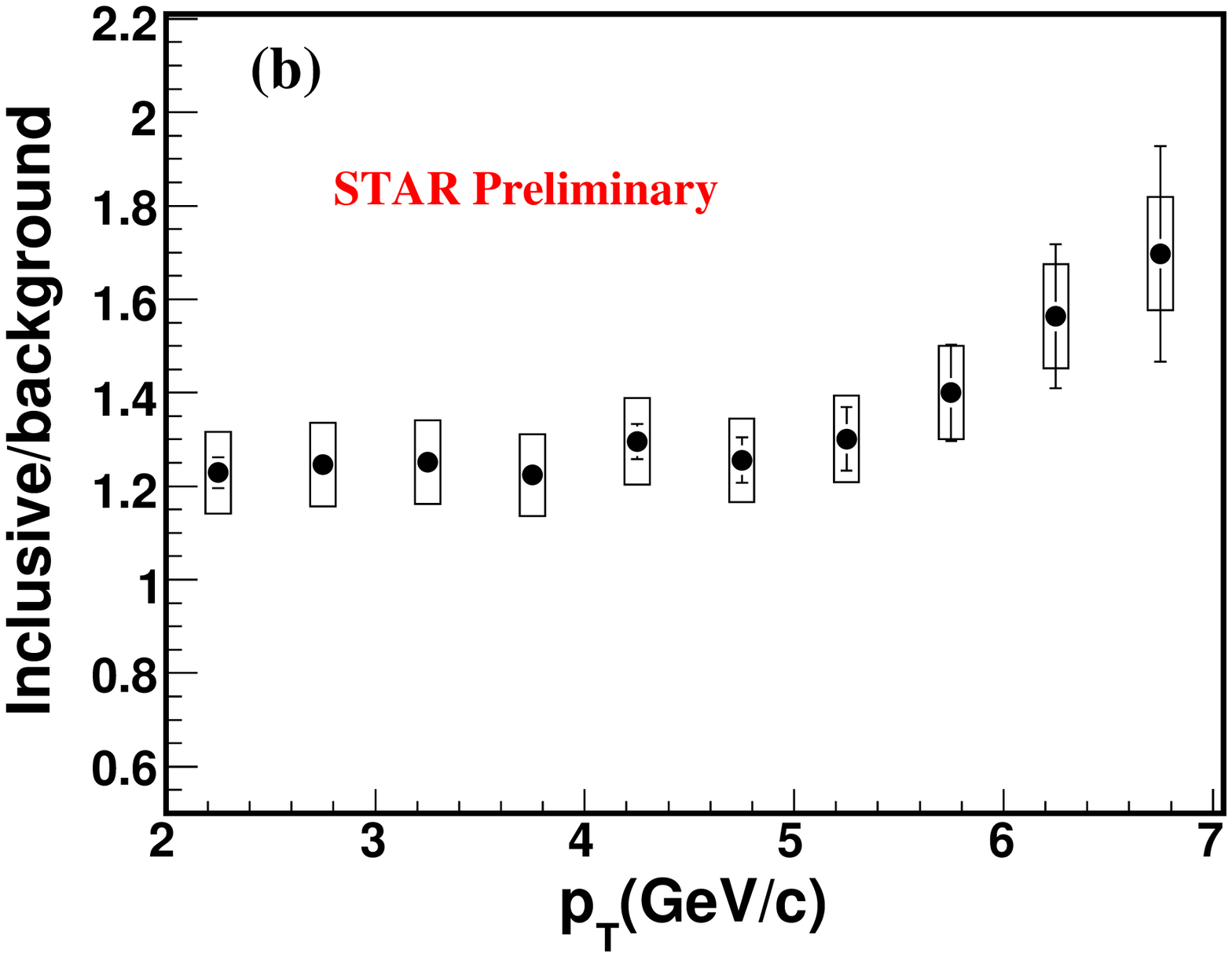,width=6.5cm}}
\vspace*{8pt} \vspace{-0.3cm} \caption{(color online) (a)Invariant
mass distributions of $e^{+}e^{-}$ pairs ($OppSign$, grey filled
area) and combinatorial background ($SameSign$, red solid line) in
p+p collisions. (b)The ratio of inclusive electron to photonic
background as a function of $p_{T}$ in p+p collisions.}\label{fig3}
\vspace{-0.3cm}
\end{figure}

\begin{figure}[th]
\centerline{\psfig{file=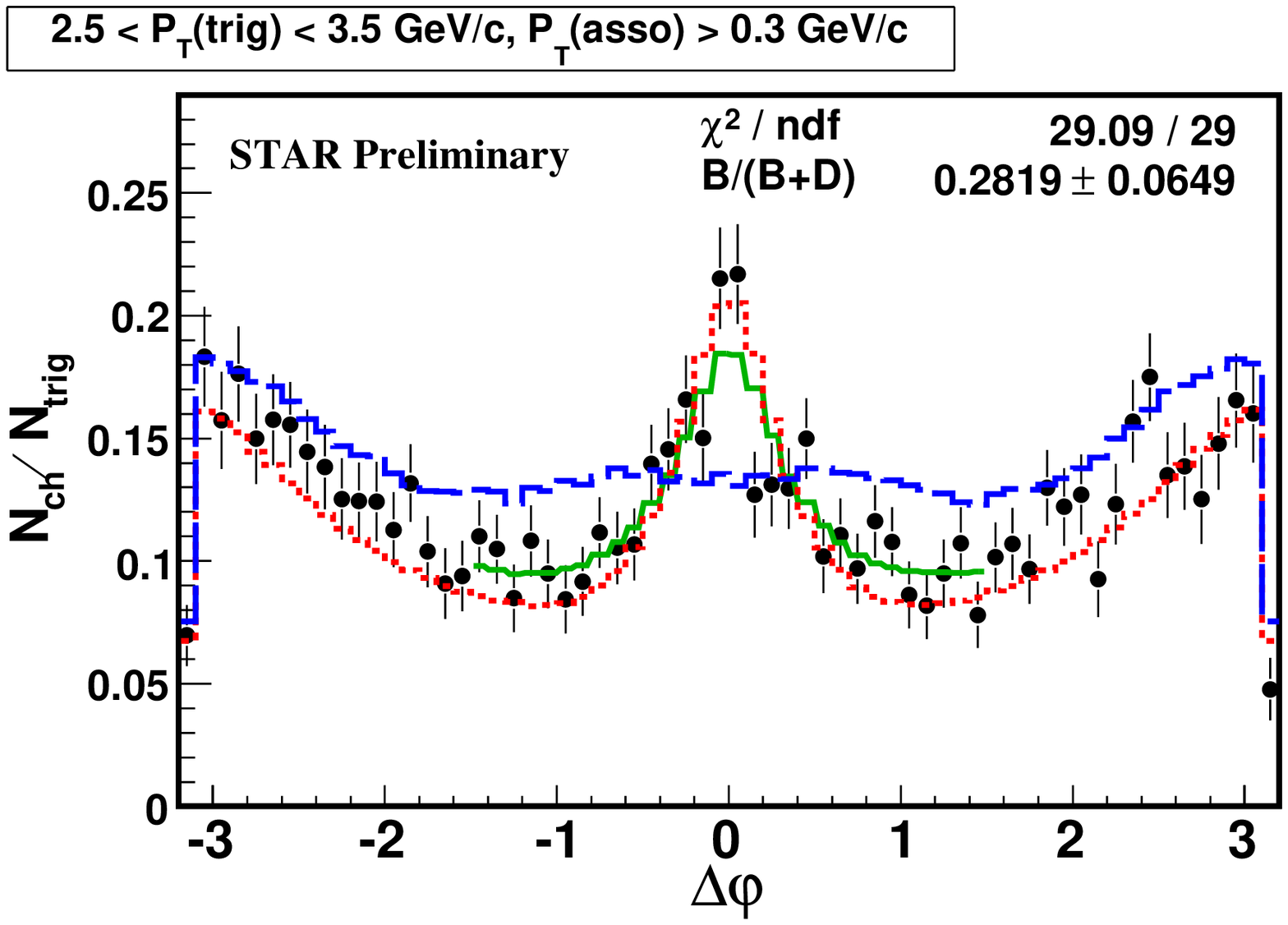,width=6.5cm}
\psfig{file=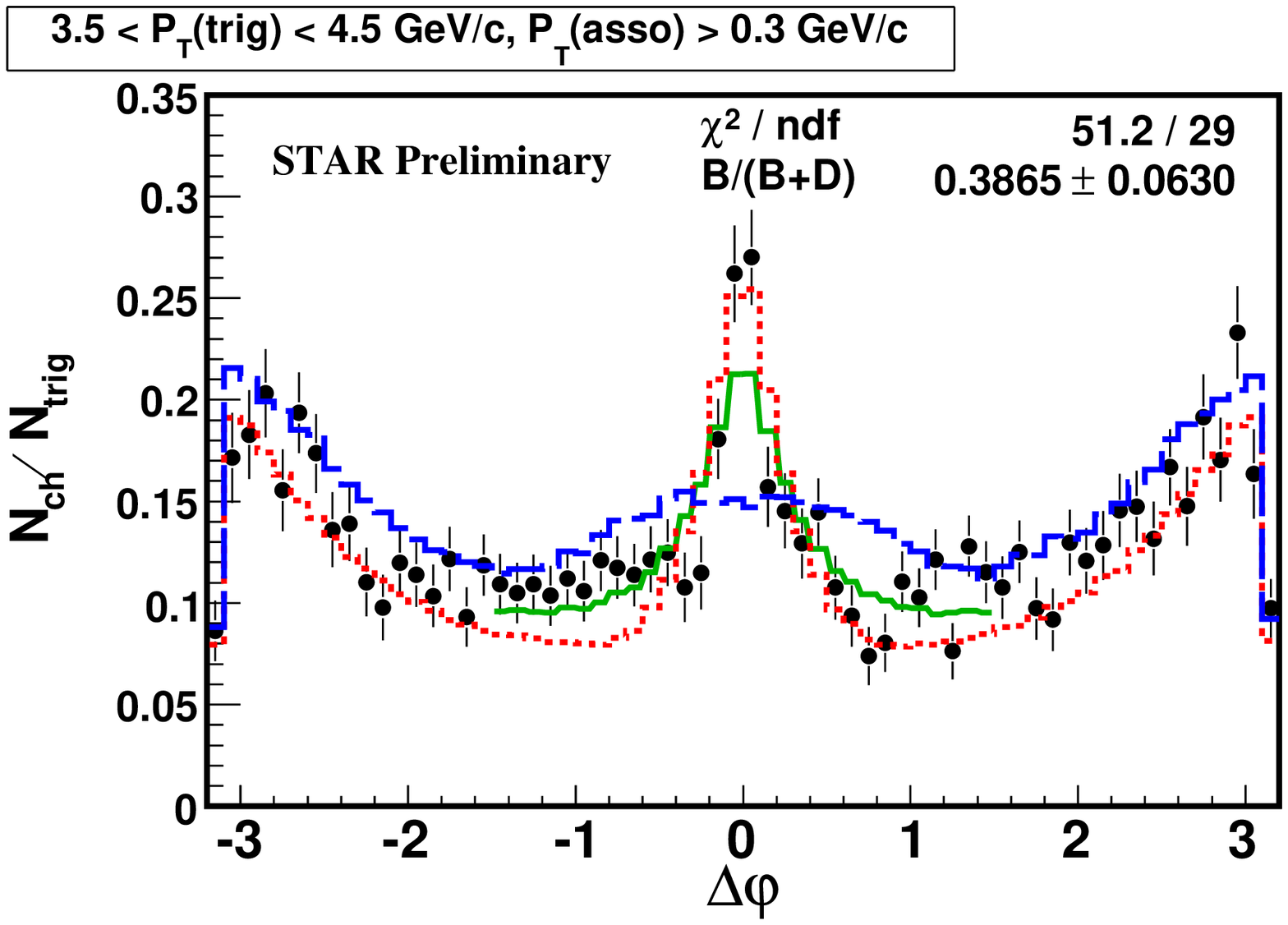,width=6.5cm}}
\centerline{\psfig{file=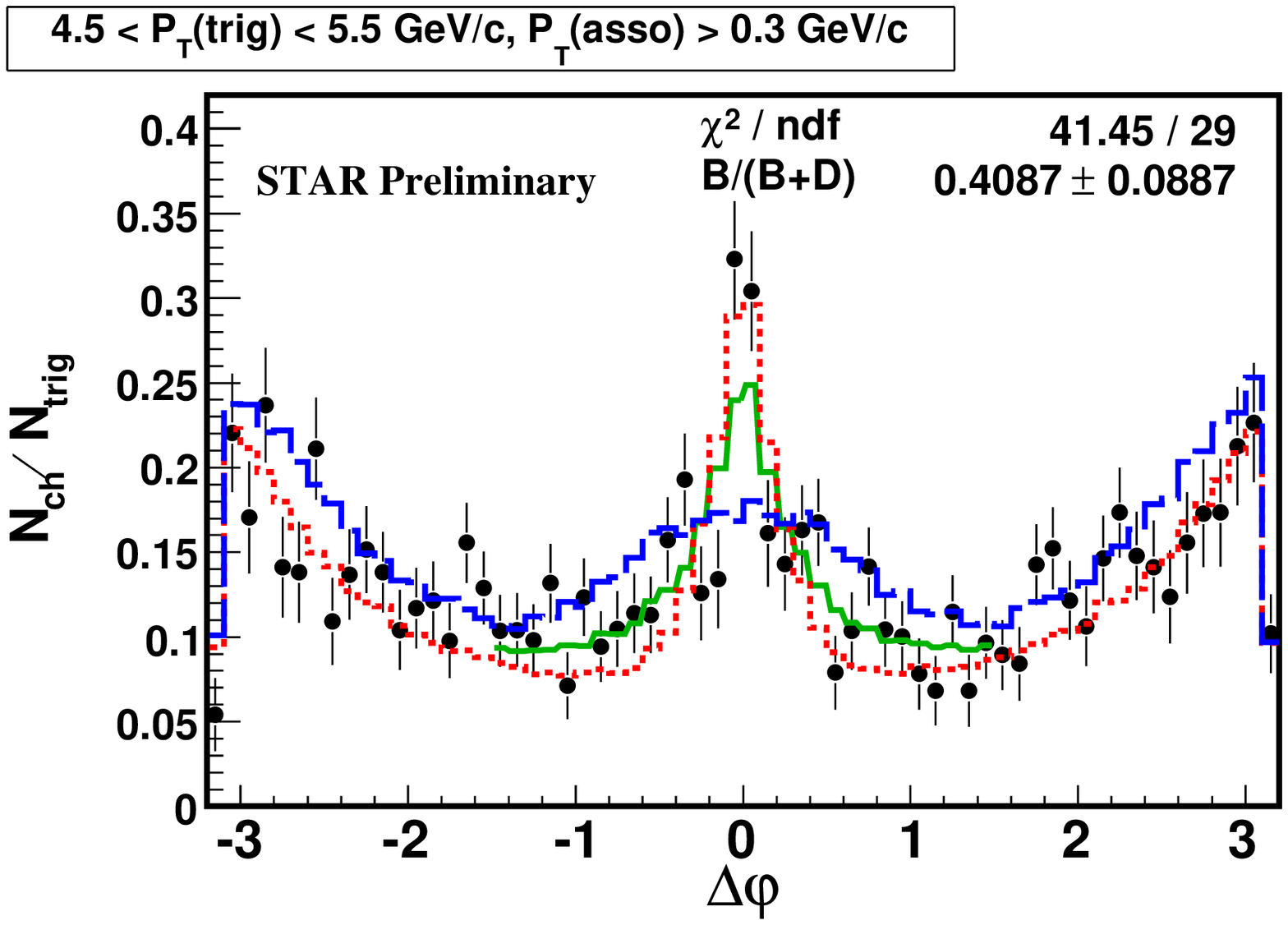,width=6.5cm}
\psfig{file=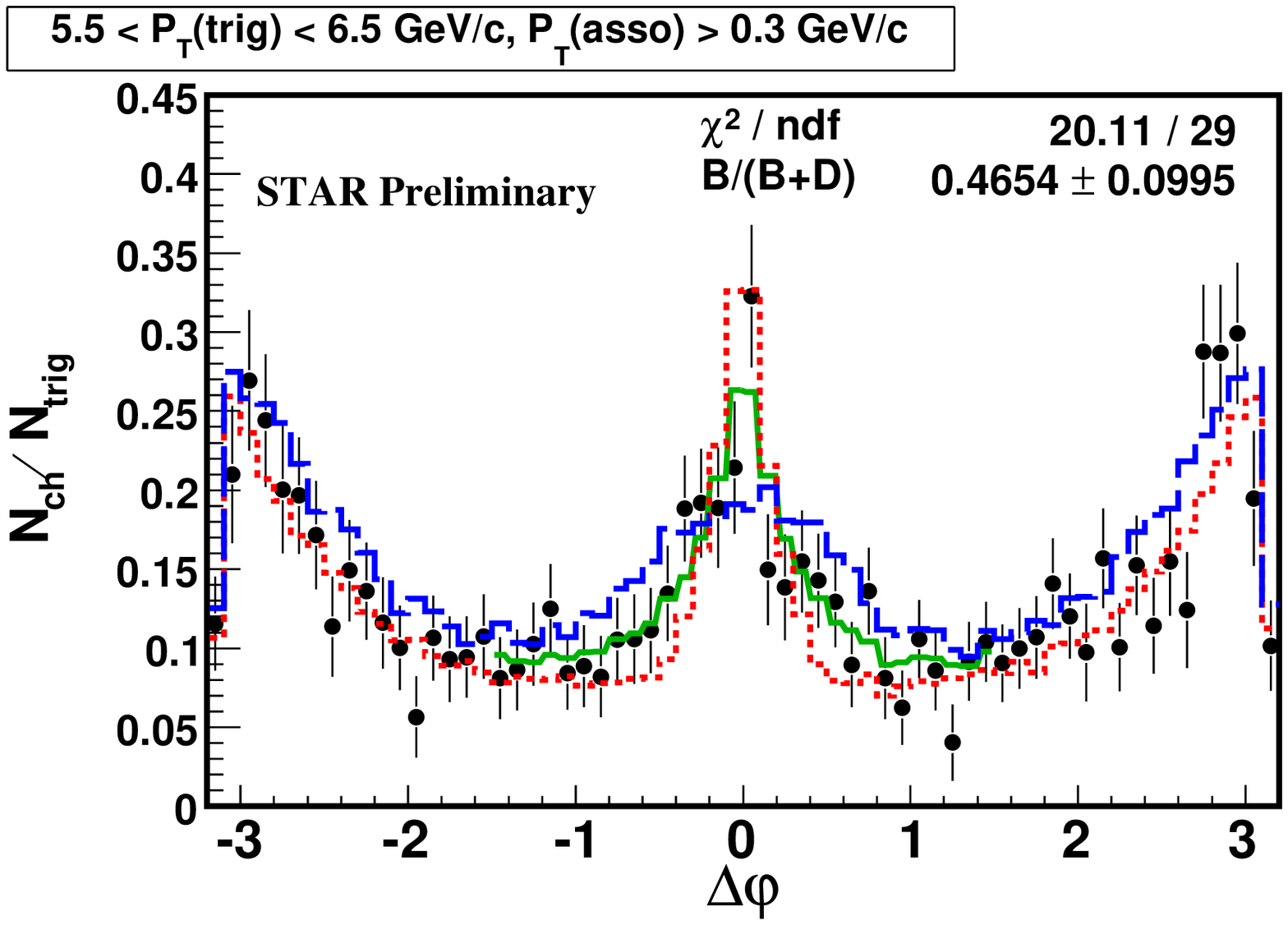,width=6.5cm}}
\vspace*{8pt} \vspace{-0.3cm} \caption{(color online)
$\Delta\varphi_{non-pho}$ distributions and the comparison to PYTHIA
simulations for four electron trigger cuts with associated hadron
$p_{T}(assoc)
>$ 0.3 GeV/c. The data are shown as dots, the simulations are depicted by blue dashed lines for $B$ meson
decays and red dotted lines for $D$ meson decays. The green solid
curves are the fits to data points using PYTHIA curves.}\label{fig4}
\vspace{-0.3cm}
\end{figure}

The left panel of Fig.~\ref{fig1} shows the $\Delta\varphi$
distributions between non-photonic electrons and inclusive charged
hadrons with six electron trigger $p_{T}$ cuts and the associated
hadron $p_{T} > 0.3$ GeV/c. The distributions are scaled by the
number of electron triggers. The solid curves and dashed curves are
for $B$ decays and for $D$ decays, respectively. There is a
significant difference between $B$ and $D$ meson decays in the
near-side correlations. The width of near-side peak for electrons
from $D$ decays is much narrower than those for the $B$ decays. The
wide width from $B$ meson decays is due to the larger energy release
($Q$ value) in the $B$ meson semi-leptonic decays leading to a broad
angular correlation between daughter hadrons and electrons in the
laboratory frame. For an electron at high $p_{T}$ from $B$ meson
decays, the $B$ meson does not have to be at high transverse
momentum because the electron can get large momentum from the
$b$-quark mass. In the case of $D$ meson decays, the $D$ meson needs
to have a large momentum in order to boost the daughter electron to
a high $p_{T}$. Fig.~\ref{fig2} shows the distributions of electron
$p_{T}$ versus its parent $p_{T}$. We found the difference in the
near-side correlations between $D$ decays and $B$ decays is largely
due to the decay kinematics, not the production dynamics. Variations
on fragmentation function from default Peterson function to $\delta$
function and on the parameter PARP(67) from $1$ to $4$ do not change
the shape of the correlations in a significant way. This difference
can help us estimate the relative $B$ and $D$ contributions to the
yields for non-photonic electrons. We will use the $\Delta\varphi$
distributions from $B$ decays and $D$ decays in PYTHIA to fit the
$\Delta\varphi$ distributions from the real data and thus measure
the $B$ contribution, $B/(B+D)$.

We further studied the particle production within a cone around
triggered high $p_{T}$ electrons from heavy quark decays. We focused
on the scalar summed $p_{T}$ distributions of inclusive charged
hadrons in the cone ($p_{T}$ refers to the transverse momentum in
the laboratory frame). Here the cone is defined by $|\eta_{h} -
\eta_{e}| < 0.35$ and $|\varphi_{h} - \varphi_{e}| < 0.35$ ($\eta$
is pseudorapidity and $\varphi$ is azimuthal angle). The summed
$p_{T}$ distributions of inclusive charged hadrons in three
triggered electron $p_{T}$ ranges are shown in the right panel of
Fig.~\ref{fig1}. The distributions are scaled to unity. The dashed
lines are for $D$ decays and the solid lines are for $B$ decays. We
also can see that there is a significant difference between $B$
decays and $D$ decays. The summed $p_{T}$ distributions for $D$
meson decays are much wider than those for $B$ meson decays. This
difference can also be used to distinguish $B$ and $D$ decay
contributions. It will be more valuable for the small acceptance
experiments to investigate the $B$ decay contribution using summed
$p_{T}$ distributions.

\section{STAR Data Analysis}

The data used in this analysis is p+p events at $\sqrt{s_{NN}} =
200$ GeV recorded by STAR in RUN
\uppercase\expandafter{\romannumeral5}. The main detectors utilized
in this analysis are the STAR Time Projection Chamber
(TPC)\cite{tpc} and the STAR Barrel Electromagnetic Calorimeter
(BEMC) with the Shower Maximum Detector (SMD)\cite{emc}. To obtain
sufficient statistics at high-$p_{T}$, we used high tower triggers
corresponding to an energy deposition of at least 2.6 GeV (HT1) and
3.5 GeV (HT2) in a single tower of the BEMC. Around 2.4 million HT1
events and 1.7 million HT2 events were used for the results
presented in this paper.

\begin{wrapfigure}{l}{0.55\textwidth}
\vspace{-0.5cm}
\psfig{file=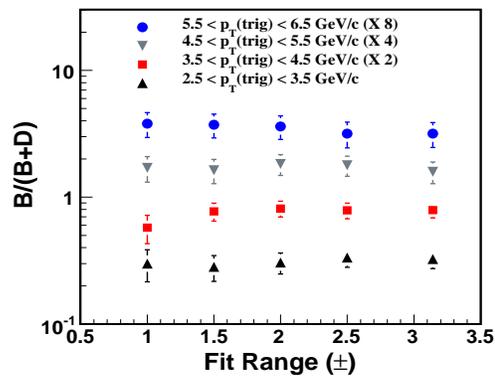,width=7.0cm,height=5.5cm}
\vspace{-0.5cm} \caption{(color online) The $B/(B+D)$ ratio as a
function of the fit range in $\Delta\varphi$. Scale factors are used
in order to separate the data points
well.}\label{fig5}\vspace{-0.3cm}
\end{wrapfigure}

Electron identification was carried out by combining ionization
energy loss in the TPC with energy deposition in the EMC and shower
profile in the SMD. The measurement of the ionization energy loss,
$dE/dx$, for charged tracks in the TPC gas is used to identify
electrons in the first stage. Requiring the $dE/dx$ values of the
selected tracks to be near the expected electron band in the region
$p_{T} >$ 2.0 GeV/c rejects a significant fraction of the hadron
background. After extrapolating the TPC tracks to the BEMC, we
require the ratio $p/E$ to be less than 1.5 using the momentum
information from the TPC, $p$, and the tower energy information from
the BEMC, $E$. Electrons will deposit almost all of their energy in
the BEMC while this is not true for hadrons. Further hadron
rejection is provided by the shower maximum detector (SMD), which
allows us to cut on the shower size with high spatial resolution. We
require the profile of the electro-magnetic shower to be within the
expectation for electrons. Combining the power of TPC and BEMC, we
can achieve an inclusive electron sample with purity $> 98\%$ in the
$p_{T}$ region up to 6.5 GeV/c. More detailed descriptions of the
electron reconstruction techniques can be found in
References.~\cite{abelev,dong}

\begin{figure}[th]
\vspace{-0.3cm}
\centerline{\psfig{file=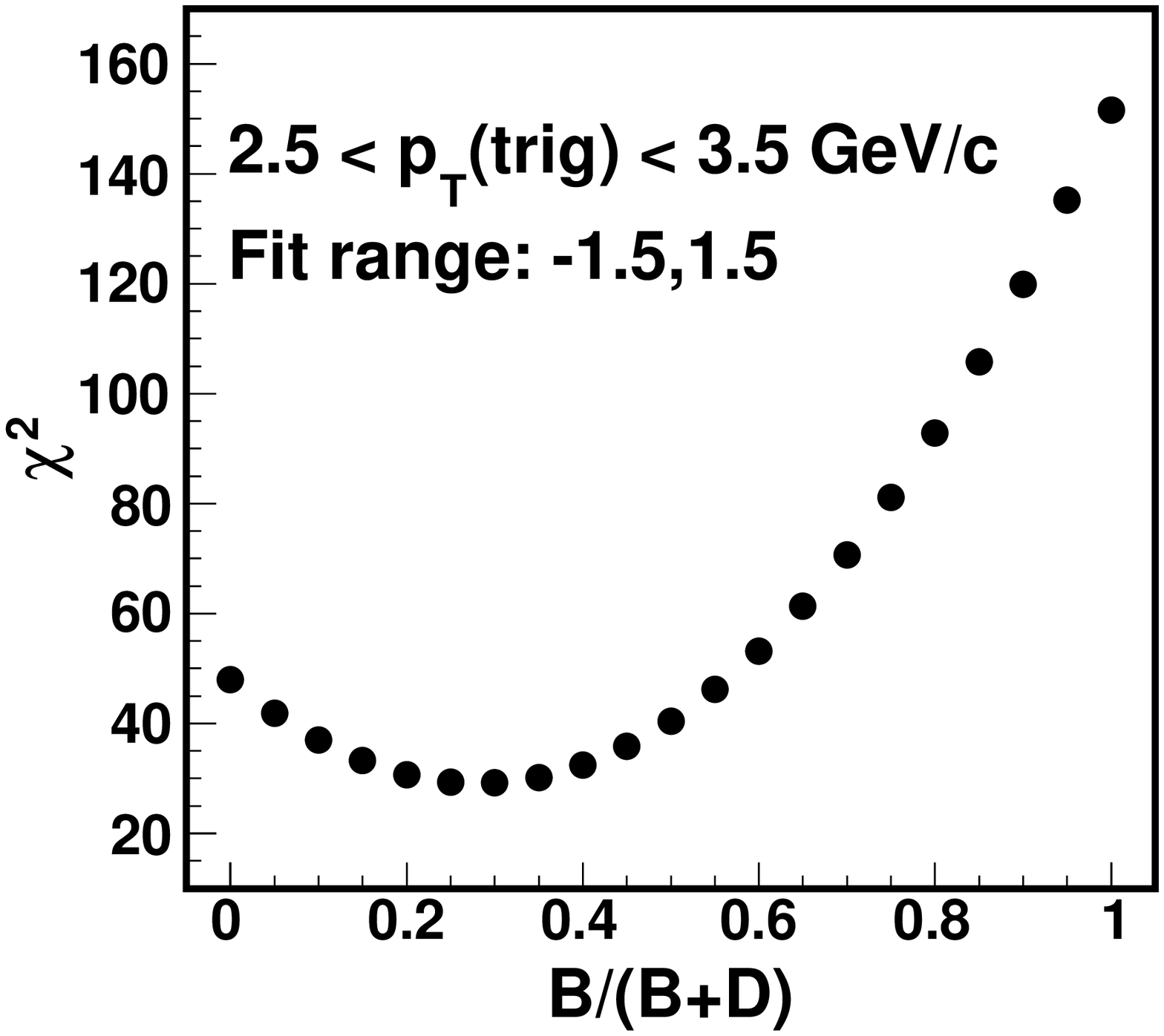,width=4.5cm}
\psfig{file=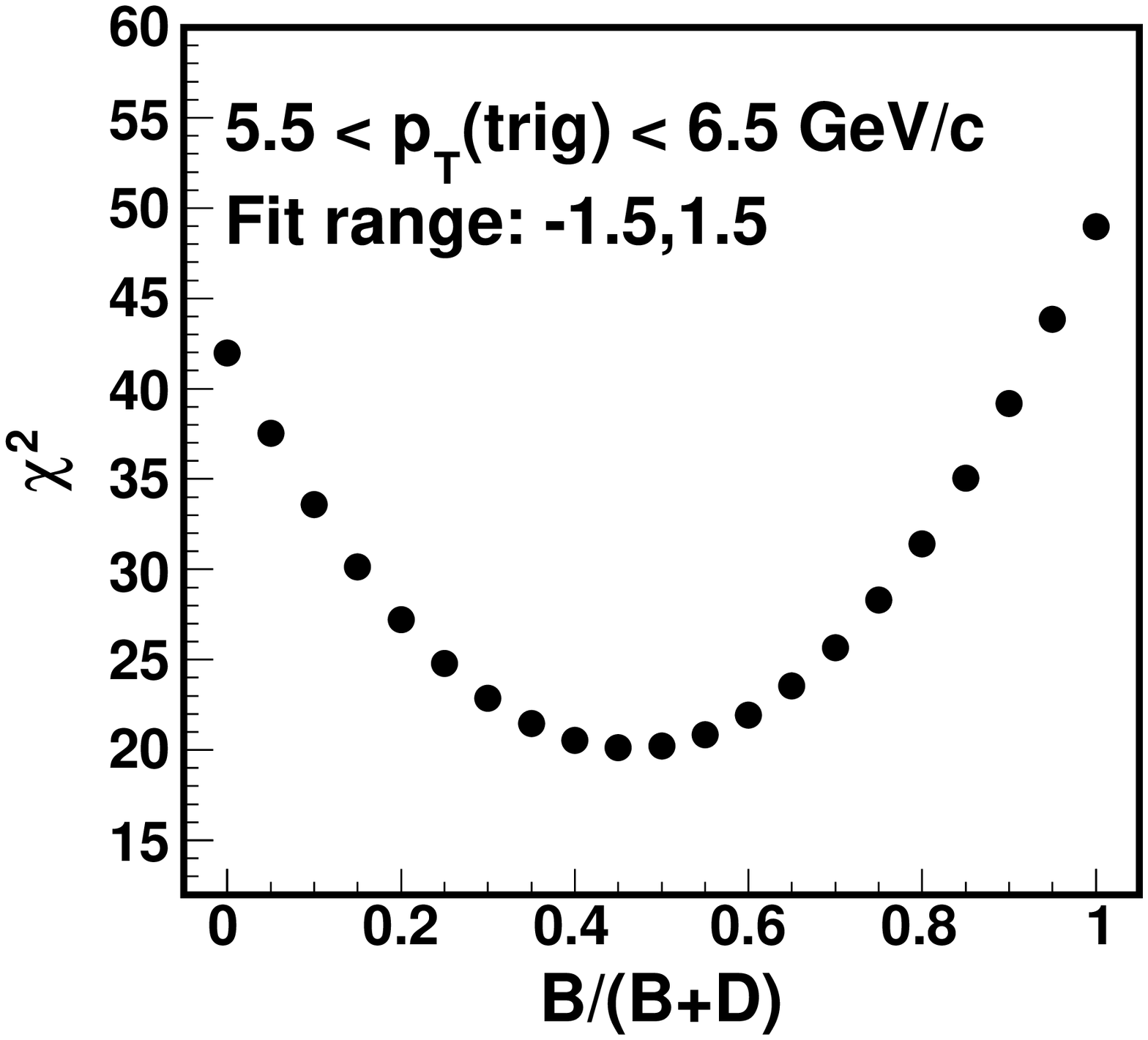,width=4.5cm}}
\vspace*{8pt} \vspace{-0.3cm}\caption{The fit $\chi^{2}$ as a
function of the $B/(B+D)$ ratio.}\label{fig6}\vspace{-0.5cm}
\end{figure}

The dominant photonic electron background is from photon conversions
and $\pi^{0}$, $\eta$ Dalitz decays, whose $e^{+}e^{-}$ pairs have
small invariant masses. We combine the electron candidates with
tracks passing a very loose cut on $dE/dx$ around the electron band.
The invariant mass distribution of $e^{+}e^{-}$ pairs ($OppSign$) is
depicted by the grey filled area in panel (a) of Fig.~\ref{fig3}.
The $OppSign$ contains the true photonic background as well as the
combinatorial background, where non-photonic electrons may be
falsely identified as photonic electrons. The combinatorial
background, which is small in p+p collisions, can be estimated by
calculating the invariant mass of same-sign electron pairs
($SameSign$) shown as red solid line in panel (a) of
Fig.~\ref{fig3}. A cut of mass $< 0.1$ GeV$/c^{2}$ rejects most of
photonic background. Panel (b) of Fig.~\ref{fig3} shows the ratio of
inclusive electron to photonic background as a function of $p_{T}$.
The bars (boxes) on the data indicate the size of statistical
(systematic) errors. A significant excess of electrons with respect
to the background has been observed, where the excess electrons are
mostly from heavy quark semi-leptonic decays. In RUN
\uppercase\expandafter{\romannumeral5} there was a larger amount of
material producing photon conversions in the STAR experimental
configuration. This led to inclusive electron to photonic background
ratios from RUN \uppercase\expandafter{\romannumeral5}
systematically lower than those from RUN
\uppercase\expandafter{\romannumeral3}
(\uppercase\expandafter{\romannumeral4}).\cite{abelev}

In order to extract the angular correlation between non-photonic
electrons and charged hadrons, we start with the semi-inclusive
electron sample. We remove the $OppSign$ sample after the mass cut
from the inclusive electron sample. The remaining electrons form the
"semi-inclusive" electron sample. The relationship of these samples
is: {\it semi-inc} $=$ {\it inc} - {\it OppSign with the mass cut}
$=$ {\it inc} $-$ ({\it reco-pho} $+$ {\it combinatorics}) $=$ {\it
inc} $-$ ({\it pho} $-$ {\it not-reco-pho} $+$ {\it combinatorics})
$=$ {\it non-pho} $+$ {\it not-reco-pho} $-$ {\it combinatorics}.
Therefore the signal can be obtained by the equation:
$\Delta\varphi_{non-pho} = \Delta\varphi_{semi-inc} -
\Delta\varphi_{not-reco-pho} + \Delta\varphi_{combinatorics}$.
$\Delta\varphi_{not-reco-pho}$ can be calculated using
$\Delta\varphi_{reco-pho}$ by an efficiency correction after
removing the photonic partner of the reconstructed-photonic
electron. The reason to remove the photonic partner is that for the
reconstructed-photonic electron the photonic partner is found while
for not-reconstructed-photonic electron the partner is missing. The
resulting e-h correlations for reconstructed photonic electrons and
not reconstructed photonic electrons cannot be related to each other
by an efficiency correction factor alone. Therefore
$\Delta\varphi_{not-reco-pho}$ can be obtained by the equation:
$\Delta\varphi_{not-reco-pho} = (1/\varepsilon -
1)\Delta\varphi_{reco-pho-no-partner}$, where $\varepsilon$ is the
photonic electron reconstruction efficiency and
$\Delta\varphi_{reco\mbox{-}pho\mbox{-}no\mbox{-}partner}$ means
reconstructed photonic electron azimuthal correlations with charged
hadrons after removing the photonic partners. The corresponding
efficiency can be calculated from simulations and is $\sim 70\%$.

\section{Results and Discussions}

\begin{wrapfigure}{l}{0.55\textwidth}
\vspace{-0.5cm}
\psfig{file=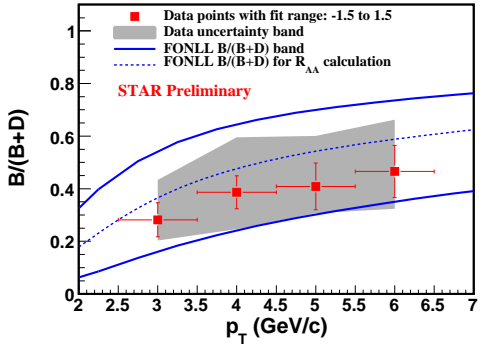,width=7.0cm}
\vspace{-0.5cm} \caption{(color online) The relative $B$
contribution to non-photonic electrons as a function of electron
$p_{T}$.}\label{fig7}\vspace{-0.3cm}
\end{wrapfigure}

Fig.~\ref{fig4} shows the $\Delta\varphi_{non-pho}$ distributions
and the comparison to PYTHIA simulations for four electron trigger
cuts with associated hadron $p_{T}(assoc)
>$ 0.3 GeV/c. The data are shown as dots, the blue dashed curves and the
red dotted curves are from PYTHIA simulations for $B$ decays and for
$D$ decays, respectively. We use PYTHIA curves to fit the data
points with the $B$ contribution as a parameter in the fit function:
$\Delta\varphi_{exp} = R\times \Delta\varphi_{B} + (1-R)\times
\Delta\varphi_{D}$, where $R$ is the $B/(B+D)$ ratio. The fits are
shown as green solid curves in Fig.~\ref{fig4}. The fit results are
consistent within statistical errors when we vary the fit range in
$\Delta\varphi$ from $\pm 1$ to $\pm \pi$ as shown in
Fig.~\ref{fig5}. As a cross check, we fixed the $B/(B+D)$ ratio to
see how the fit $\chi^{2}$ changes. Fig.~\ref{fig6} shows the fit
$\chi^{2}$ as a function of the $B/(B+D)$ ratio with the fit range
in $\Delta\varphi$ from $-1.5$ to $1.5$ for two trigger $p_{T}$
cuts. The $\chi^{2}$ is sensitive to the $B/(B+D)$ ratio, and so is
it for other trigger $p_{T}$ cuts and for other fit ranges. As a
systematic check, we allowed an overall normalization factor in the
fit function to float. It gives a normalization factor close to
unity, and consistent $B/(B+D)$ ratios.


The $B$ semi-leptonic decay contribution to non-photonic electrons
as a function of $p_{T}$ is shown in Fig.~\ref{fig7}. The bars show
the size of statistical errors. The grey band indicates the data
uncertainties including statistical errors, and systematic
uncertainties introduced by photonic electron reconstruction
efficiency (dominant), different fit ranges and different fit
functions. The blue solid curves show the range of relative bottom
contribution from the Fixed Order Next-to-Leading Log (FONLL)
calculations.\cite{cacciari} The dashed line is the $B/(B+D)$ ratio
in FONLL used for the default non-photonic electron $R_{AA}$
calculation. The preliminary STAR data are consistent with the FONLL
calculation. Taking the observed non-photonic electron suppression
in central Au+Au collisions at RHIC together with the measured
$B/(B+D)$ ratio in p+p collisions suggests that $B$ mesons may be
suppressed in dense medium. A calculation from Adil and Vitev
predicts suppression of $B$ mesons comparable to that of $D$ mesons
at transverse momentum as low as $p_{T} \sim 10$ GeV/c.\cite{adil2}

\section{Conclusion}
In conclusion, correlations of non-photonic electrons with charged
hadrons are studied using the PYTHIA model in 200 GeV p+p
collisions. The measurement of azimuthal correlations between
non-photonic electrons and charged hadrons in p+p collisions at
$\sqrt{s_{NN}}=200$ GeV from STAR has been presented. The
experimental results are compared with PYTHIA simulations to measure
the relative $B$ contribution to non-photonic electrons. Within the
present statistical and systematic errors, the preliminary data
analysis based on PYTHIA model indicates at $p_{T} \sim 4.0 - 6.0$
GeV/c the measured $B$ contribution to non-photonic electrons is
comparable to the $D$ contribution and that the measured $B/(B+D)$
ratios are consistent with the FONLL theoretical calculation.
Together with the observed suppression of non-photonic electrons in
central Au+Au collisions, the measured $B/(B+D)$ ratios imply that
the bottom quark may be suppressed in central Au+Au collisions at
RHIC.

\vspace{0.5cm} \noindent This work is partially supported by NSFC
under the Grant No. 10575042, 10610285.

\end{document}